\documentclass[prd,aps,twocolumn,showpacs,nofootinbib]{revtex4-1}

\usepackage{amsfonts,color,graphicx,epsfig,amsmath,multirow}

\begin{document}

\title{The $\rho$-meson longitudinal leading-twist distribution amplitude}
\author{Hai-Bing Fu}
\author{Xing-Gang Wu}
\email{Email: wuxg@cqu.edu.cn}
\author{Hua-Yong Han}
\author{Yang Ma}
\author{Huan-Yu Bi}
\address{Department of Physics, Chongqing University, Chongqing 401331, P.R. China}

\begin{abstract}
In the present paper, we suggest a convenient model for the vector $\rho$-meson longitudinal leading-twist distribution amplitude $\phi_{2;\rho}^\|$, whose distribution is controlled by a single parameter $B^\|_{2;\rho}$. By choosing proper chiral current in the correlator, we obtain new light-cone sum rules (LCSR) for the $B\to\rho$ TFFs $A_1$, $A_2$ and $V$, in which the $\delta^1$-order $\phi_{2;\rho}^\|$ provides dominant contributions. Then we make a detailed discussion on the $\phi_{2;\rho}^\|$ properties via those $B\to\rho$ TFFs. A proper choice of $B^\|_{2;\rho}$ can make all the TFFs agree with the lattice QCD predictions. A prediction of $|V_{\rm ub}|$ has also been presented by using the extrapolated TFFs, which indicates that a larger $B^{\|}_{2;\rho}$ leads to a larger $|V_{\rm ub}|$. To compare with the BABAR data on $|V_{\rm ub}|$, the longitudinal leading-twist DA $\phi_{2;\rho}^\|$ prefers a doubly-humped behavior.


\end{abstract}

\maketitle

\section{Introduction}

The meson light-cone distribution amplitude (LCDA) relates to the matrix elements of the nonlocal light-ray operators sandwiched between the hadronic state and the vacuum. It provides underlying links between the hadronic phenomena at the small and large distances. The leading-twist (twist-2) LCDA arouses people's great interests due to its role in hard exclusive processes~\cite{lepage}, i.e. it always provides dominant contributions to those processes. The investigation on the LCDAs has been the subject of numerous studies, especially for the simpler LCDAs of the pseudoscalar mesons pion, kaon and etc. As for the vector $\rho$ meson, its LCDAs are much more involved due to its complex structures: there are chiral-even or chiral-odd LCDAs for the $\rho$ meson, the $\rho$ meson has longitudinal ($\|$) and transverse ($\perp$) polarization states. Several approaches have been adopted to study the $\rho$-meson LCDA properties, such as the QCD sum rules~\cite{rho1,rho2,rho3,rho6}, the light-cone quark model~\cite{rho4}, the AdS/QCD model~\cite{AdS:3} and etc. The properties of the $\rho$-meson LCDA are also helpful for us to understand other vector mesons' LCDAs. That is, by taking the ${\rm SU}_f(3)$-breaking effect into consideration, one can further study the $K^*$ meson LCDAs~\cite{rho4,rho7,rho8}.

\begin{table}[htb]
\centering
\begin{tabular}{ | c | c| c | c | c | }
\hline
 twist & ~~$\delta^0$~~  & ~~$\delta^1$~~ & ~~$\delta^2$~~ & ~~$\delta^3$~~   \\
\hline
~~2~~  & $\phi_{2;\rho}^\bot$ & $\phi_{2;\rho}^\|$ & / & / \\
\hline
~~3~~  & / & $\phi_{3;\rho}^\bot$,$\psi_{3;\rho}^\bot$,$\Phi_{3;\rho}^\|$,$\widetilde\Phi_{3;\rho}^\|$ & $\phi_{3;\rho}^\|$,$\psi_{3;\rho}^\|$,$\Phi_{3;\rho}^\bot$ & / \\
\hline
~~4~~  & / & / & $\phi_{4;\rho}^\bot$,$\psi_{4;\rho}^\bot$,$\Psi_{4;\rho}^\bot$,$\widetilde{\Psi} _{4;\rho}^\bot$ & $\phi_{4;\rho}^\|$,$\psi_{4;\rho}^\|$ \\
\hline
\end{tabular}
\caption{The $\rho$-meson LCDAs up to twist-4 and $\delta^3$-order.} \label{DA_delta}
\end{table}

To understand the complex structures of the $\rho$-meson LCDAs, as suggested in Refs.\cite{Ball04:Brho,P.Ball:1998-1999}, it is convenient to arrange them by a parameter $\delta$, i.e. $\delta \simeq m_\rho/m_b\sim 16\%$. A classification of the $\rho$-meson twist-2, twist-3 and twist-4 LCDAs up to $\delta^3$-order have been collected in Ref.~\cite{fuhb2014:Brho}, which are rearranged in Table \ref{DA_delta}. Up to twist-4, there are fifteen $\rho$-meson LCDAs. All those $\rho$-meson LCDAs, especially the higher-twist DAs, are far from affirmation, thus, it is helpful to find a reliable way to study the properties of a specific $\rho$-meson LCDA. In the paper, we shall concentrate our attention on the $\rho$-meson longitudinal leading-twist distribution amplitude $\phi_{2;\rho}^\|$, which is at the $\delta^{1}$-order. For the purpose, we adopt the $B\to\rho$ transition form factors (TFFs) to study the $\phi_{2;\rho}^\|$ properties.

The $B\to\rho$ TFFs are key components for the semileptonic decay $B\to\rho l\nu$. When the leptons are massless, only three TFFs $A_1$, $A_2$ and $V$ provide non-zero contributions to the $B\to\rho l\nu$ decay~\cite{Ball1997:Brho}. In the following, we shall adopt the QCD light-cone sum rules (LCSR)~\cite{LCSR:1,LCSR:2,LCSR:3} to deal with those non-zero $B\to\rho$ TFFs. In comparison to the convention SVZ sum rules~\cite{svz}, the LCSR is based on the operator product expansion near the light cone, and all its non-perturbative dynamics are parameterized into LCDAs of increasing twists. One advantage of LCSR lies in that it allows to incorporate information about high-energy asymptotics of correlation functions in QCD, which is accumulated in the LCDAs. Furthermore, as a tricky point of the LCSR approach, by proper choosing the correlation function (correlator), especially by choosing the chiral correlator, one can highlight the wanted LCDAs' contributions while highly suppress the unwanted LCDAs' contributions to the LCSR~\cite{chiral1,chiral2,chiral3}. As shall be shown later, by taking a proper chiral correlator, we can highlight the contributions of $\phi_{2;\rho}^\|$ to the $B\to\rho$ TFFs. The obtained LCSRs for those TFFs shall show strong dependence on $\phi_{2;\rho}^\|$. Thus, via the comparison with the data or predictions from other approaches, it shall provides us good opportunities to determine the behavior of $\phi_{2;\rho}^\|$.

The remaining parts of the paper are organized as follows. In Sec. II, we present the formulas for the $B\to\rho$ TFFs under the LCSR approach. In Sec. III, we present our numerical results and discussions on the $B\to\rho$ TFFs, the $B\to\rho$ semi-leptonic decay width and the Cabibbo-Kobayashi-Maskawa (CKM) matrix element $|V_{\rm ub}|$. Sec. IV is reserved for a summary.

\section{Calculation technology}

To derive the LCSRs for the $B\to\rho$ TFFs, we suggest to start from the following chiral correlator
\begin{eqnarray}
&&\Pi _\mu(p,q)= \nonumber\\
&& i\int d^4x e^{iq\cdot x}\langle\rho (p,\lambda)|{\rm T} \left\{\bar q_1(x)\gamma_\mu(1-\gamma_5) b(x), j_B^\dag (0)\right\} |0\rangle, \label{correlator}
\end{eqnarray}
where the current $j_B^\dag (x)=i \bar b(x)(1 - \gamma_5)q_2(x)$. This is different from our previous choice of $j_B^\dag (x)=i \bar b(x)(1 + \gamma_5)q_2(x)$~\cite{fuhb2014:Brho}, in which the $\delta^0$-order $\phi_{2;\rho}^\perp$ provide dominant contribution. The advantage of our present new choice lies in that one can highlight the contributions from the chiral-even $\rho$-meson LCDAs such as $\phi_{2;\rho}^\|$, $\phi_{3;\rho}^\bot$, $\psi_{3;\rho}^\bot$, $\Phi_{3;\rho}^\|$, $\widetilde \Phi_{3;\rho}^\|$, $\phi_{4;\rho}^\|$ and $\psi_{4;\rho}^\|$; while the chiral-odd LCDAs $\phi_{2;\rho}^\bot$, $\phi_{3;\rho}^\|$, $\psi_{3;\rho}^\|$, $\Phi_{3;\rho}^\bot$, $\phi_{4;\rho}^\bot$, $\psi_{4;\rho}^\bot$, $\Psi_{4;\rho}^\bot$, $\widetilde{\Psi} _{4;\rho}^\bot$ provide zero contributions. We also find that within the remaining chiral-even LCDAs, only the LCDAs $\phi_{2;\rho}^\|$, $\phi_{3;\rho}^\bot$ and $\psi_{3;\rho}^\bot$ provide dominant contributions to the LCSR, while the remaining LCDAs as $\phi_{4;\rho}^\|$, $\psi_{4;\rho}^\|$, $\Phi_{3;\rho}^\|$ and $\widetilde \Phi_{3;\rho}^\|$ provide negligible contribution. Moreover, the twist-3 DAs $\psi_{3;\rho}^\bot$ and $\phi_{3;\rho}^\bot$ (and the later simplified $A_\rho^\| (u) = \int_0^u dv [\phi _{2;\rho }^\| (v) - \phi _{3;\rho }^ \bot (v)]$) can be related to $\phi_{2;\rho}^\|$ under the Wandzura-Wilczek approximation~\cite{WW,Ball1997:Brho}
\begin{eqnarray}
\phi_{3;\rho}^\bot(x,\mu ) &=& \left[\Phi_\rho^{0-x}(x,\mu) + \Phi_\rho^{x-1}(x,\mu) \right]/2, \\
\psi_{3;\rho}^\bot(x,\mu ) &=& 2\left[ \bar x \Phi_\rho^{0-x}(x,\mu) + x \Phi_\rho^{x-1}(x,\mu) \right], \\
A_\rho^\|(x,\mu) &=& \left[ \bar x \Phi_\rho^{0-x}(x,\mu) + x \Phi_\rho^{x-1}(x,\mu) \right]/2 ,
\end{eqnarray}
where $\Phi_\rho^{0-x}(x,\mu) =  \int_0^x {dv} \frac{\phi _{2;\rho }^\| (v,\mu )}{\bar v} $ and $\Phi_\rho^{x-1}(x,\mu) = \int_x^1 dv \frac{\phi _{2;\rho }^\| (v,\mu)}{v}$. Thus, the leading-twist $\phi_{2;\rho}^\|$ provides dominant contribution to the LCSRs either directly or indirectly. Then, as required, those TFFs can indeed provide us a useful platform for testing the properties of $\phi_{2;\rho}^\|$.

The correlator (\ref{correlator}) is an analytic function of $q^2$ defined at both the space-like and the time-like $q^2$-regions. The correlator can be treated by inserting a completed set of intermediate hadronic states in physical region. It can also be treated in the framework of the operator product expansion in deep Euclidean region, then all its non-perturbative dynamics are parameterized into LCDAs. Those two results can be related by the dispersion relation, and the final LCSR can be achieved by applying the Borel transformation.

Following the standard LCSR procedures, we obtain the LCSRs for the $B\to\rho$ TFFs
\begin{widetext}
\begin{eqnarray}
f_B A_1(q^2) e^{-m_B^2/M^2} &=& \frac{2m_b^2 m_\rho f_\rho ^\|}{m_B^2  (m_B + m_\rho)} \bigg\{\int_0^1\frac{du}{u}e^{-s(u)/M^2} \left[ \Theta(c(u,s_0))\phi_{3;\rho}^\bot(u) - \frac{m_\rho^2}{u M^2}\widetilde\Theta(c(u,s_0)) C_\rho^\|(u) \right] \nonumber\\
&& - m_\rho^2\int {\cal D} \underline\alpha\int dv e^{-s(X) /M^2} \frac{1}{X^2 M^2} \Theta(c(X,s_0)) [\Phi_{3;\rho}^\|(\underline \alpha ) + {\widetilde \Phi}_{3;\rho}^\|(\underline \alpha )] \bigg\},   \label{TFF:A1}\\
f_B A_2(q^2)e^{-m_B^2/M^2} &=& \frac{m_b^2 m_\rho (m_B + m_\rho )f_\rho^\| }{m_B^2 }\bigg\{2\int_0^1 \frac{du}{u} e^{- s(u)/M^2} \bigg[\frac{1}{uM^2}\widetilde\Theta(c(u,s_0)) A_\rho^\|(u) + \frac{m_\rho^2}{u M^4}\widetilde{\widetilde\Theta}(c(u,s_0)) C_\rho^\|(u) \nonumber\\
&& \!\!\!\!\!\!\!\!\!\!\!\!\!\!\!\!+ \frac{m_b^2 m_\rho^2}{4u^4M^6} \widetilde{\widetilde{\widetilde\Theta}}(c(u,s_0)) B_\rho^\|(u) \bigg] + m_\rho^2\int {\cal D} \underline\alpha \int{dv} e^{ - s(X)/M^2} \frac{1}{X^3 M^4}\Theta(c(X,s_0)) [\Phi_{3;\rho}(\widetilde {\underline \alpha }) + \widetilde \Phi_{3;\rho}^\|(\widetilde {\underline \alpha  })] \bigg\},  \label{TFF:A2}\\
f_B V(q^2)e^{-m_B^2/M^2} &=& \frac{m_b^2 m_\rho (m_B+m_\rho)f_\rho^\| }{2m_B^2}\int_0^1 du e^{- s(u)/M^2}\frac{1}{u^2 M^2}\widetilde\Theta(c(u,s_0))\psi_{3;\rho}^ \bot(u),   \label{TFF:V}
\end{eqnarray}
\end{widetext}
where $s(\varrho)=[m_b^2-\bar \varrho(q^2-\varrho m_\rho^2)]/\varrho$ with $\bar \varrho = 1 - \varrho$ ($\varrho$ stands for $u$ or $X$), and $X=a_1 + v a_3$. $f_\rho^\|$ represents the $\rho$-meson decay constant. $c(u,s_0)=u s_0 - m_b^2 + \bar u q^2 - u \bar u m_\rho^2$. $\Theta(c(\varrho,s_0))$ is the usual step function: when $c(\varrho,s_0)<0$, it is zero; otherwise, it is $1$. $\widetilde\Theta (c(u,s_0))$ and $\widetilde{\widetilde\Theta}(c(u,s_0))$ are defined via the integration
\begin{widetext}
\begin{eqnarray}
&& \int_0^1 \frac{du}{u^2 M^2} e^{-s(u)/M^2}\widetilde\Theta(c(u,s_0))f(u)
= \int_{u_0}^1\frac{du}{u^2 M^2} e^{-s(u)/M^2}f(u) + \delta(c(u_0,s_0)),
\label{Theta1}\\
&& \int_0^1 \frac{du}{2u^3 M^4} e^{-s(u)/M^2} \widetilde{\widetilde\Theta}(c(u,s_0))f(u)
= \int_{u_0}^1 \frac{du}{2u^3 M^4} e^{-s(u)/M^2}f(u)+\Delta(c(u_0,s_0)), \label{Theta2}
\end{eqnarray}
where $\delta(c(u,s_0)) = e^{-s_0/M^2}\frac{f(u_0)}{{\cal C}_0}$ and
\begin{equation}
\Delta(c(u,s_0))= e^{-s_0/M^2}\bigg[\frac{1}{2 u_0 M^2}\frac{f(u_0)} {{\cal C}_0}  \left. -\frac{u_0^2}{2 {\cal C}_0} \frac{d}{du}\left( \frac{f(u)}{u{\cal C}} \right) \right|_{u = {u_0}}\bigg], \nonumber
\end{equation}
\end{widetext}
${\mathcal C}_0 = m_b^2 + {u_0^2}m_\rho ^2 - {q^2}$ and $u_0\in[0,1]$ is the solution of $c(u_0,s_0)=0$. The simplified DAs $B_\rho^\| (u) = \int_0^u dv \phi_{4;\rho}^\|(v)$ and $C_\rho^\| (u) = \int_0^u dv \int_0^v dw$ $[\psi_{4;\rho}^\|(w) + \phi _{2;\rho }^\|(w) - 2\phi_{3;\rho}^\bot(w)]$.

\section{Numerical results}

\subsection{Input parameters}

\begin{table}[htb]
\begin{center}
\begin{tabular}{c c c c c}
\hline
$B_{2;\rho}^\|$ & $A_{2;\rho}^\|({\rm GeV^{-1}})$ & $b_{2;\rho}^\|({\rm GeV}^{-1})$ &  $a_2^{\|}(1{\rm GeV})$ & $a_2^{\|}(2.2{\rm GeV})$ \\
\hline
  $-0.2$	&	 28.96 	&	0.643 	&	 $-0.180$    &	$-0.152$\\
  $-0.1$	&	 27.88 	&	0.628 	&	 $-0.080$    &	$-0.067$\\
  $~~0.0$	&	 26.24 	&	0.604 	&	 $+0.026$	 &  $+0.022$\\
  $+0.1$	&	 24.15 	&	0.572 	&	 $+0.140$	 &  $+0.118$\\
  $+0.2$	&	 21.91 	&	0.537 	&	 $+0.258$	 &  $+0.217$\\
\hline
\end{tabular}
\caption{The parameters for the $\rho$-meson leading-twist LCDA $\psi_{2;\rho}^{\|}$, and its corresponding second Gegenbauer moment $a_2^{\|}$ at 1 GeV and 2.2 GeV, respectively. }   \label{DA_parameter_1}
\end{center}
\end{table}

The leading-twist LCDA $\phi^\|_{2;\rho}$ can be derived from its light-cone wavefunction (LCWF), i.e.
\begin{eqnarray}
\phi_{2;\rho}^\|(x,\mu_0) = \frac{ 2\sqrt{3}}{ \widetilde{f}_\rho^\|}\int_{|{\bf k}_\bot|^2\leq\mu^2_0}\frac{d{\bf k}_\bot}{16\pi^3}\psi_{2;\rho}^\|(x,{\bf k}_\bot)\,,
\end{eqnarray}
where $\widetilde{f}_\rho^\| = f_\rho^{\|}/\sqrt{5}$. One way of constructing the LCWF model has been suggested in by Wu and Huang~\cite{XGWu:2010}. Following the idea, the WH-DA model for $\phi _{2;\rho }^\|$ is
\begin{eqnarray}
&&\phi _{2;\rho }^\| (x,\mu_0) = \frac{{A_{2;\rho}^\| \sqrt {3x\bar x} {m_q}}}{{8{\pi ^{3/2}}\widetilde f_\rho ^\| b_{2;\rho}^\| }}[1 + {B_{2;\rho}^\| }C_2^{3/2}(\xi )] \nonumber\\
&& \times\left[ {{\rm{Erf}}\left( {b_{2;\rho}^\| \sqrt {\frac{{{\mu^2_0} + m_q^2}}{{x\bar x}}} } \right) - {\rm{Erf}}\left( {b_{2;\rho}^\| \sqrt {\frac{{m_q^2}}{{x\bar x}}} } \right)} \right], \label{DA:WH}
\end{eqnarray}
where $\mu_0$ is the factorization scale, the error function $\textrm{Erf}(x) ={2}\int^x_0 e^{-t^2} dt/{\sqrt{\pi}}$ and the constitute light-quark mass $m_q\simeq300$ MeV. Apart from the normalization condition, we take the average of the transverse momentum $\langle{\bf k}_\bot^2\rangle_{2;\rho}^{1/2}\simeq 0.37$ GeV~\cite{XHGuo1991} as another constraint. The DA parameters for $B^\|_{2;\rho}\in[-0.2,0.2]$ are presented in Table~\ref{DA_parameter_1}, in which its second Gegenbauer moment $a_2^{\|}$ under two typical scales, i.e. $\mu_0=1.0{\rm GeV}$ and $2.2{\rm GeV}$, are also presented. The value of $a_2^{\|}$ at $2.2{\rm GeV}$ is obtained by QCD evolution~\cite{lepage}. Table~\ref{DA_parameter_1} shows $B_{2;\rho}^\| \sim a^{\perp}_2$. Another way of constructing the LCWF has been suggested under the AdS/QCD-DA theory~\cite{Brodsky:2007hb},
\begin{eqnarray}
\psi_{2;\rho}^{\|}(x,\zeta)= \mathcal{N}_{\|} \frac{\kappa} {\sqrt{\pi}}\sqrt{x\bar{x}} \exp \left(-\frac{\kappa^2 \zeta^2}{2}\right) \exp\left(-\frac{m_f^2}{2\kappa^2 x\bar{x}} \right),\nonumber\\
\end{eqnarray}
which leads to the AdS/QCD-DA model~\cite{AdS:2}
\begin{eqnarray}
\phi_{2;\rho}^{\|}(x,\mu_0) &=&\frac{ 3 m_f}{\pi f_\rho^\|} \int d\zeta \mu_0 J_1(\mu_0 \zeta) \frac{\psi_{2;\rho}^\|(x,\zeta)}{x\bar x} , \label{phi_4}
\end{eqnarray}
where $\kappa = m_\rho/\sqrt{2}$, $m_f=0.14{\rm GeV}$~\cite{Soyez:2007kg, Forshaw:2006np, Forshaw:2004vv}. ${\cal N}_\|$ can be fixed by the normalization condition, e.g. we have ${\cal N}_{\|}|_{\mu_{0}=1{\rm GeV}}=1.221$ and ${\cal N}_{\|}|_{\mu_{0}=2.2{\rm GeV}}=1.219$.

\begin{figure}[h]
\begin{center}
\includegraphics[width=0.45\textwidth]{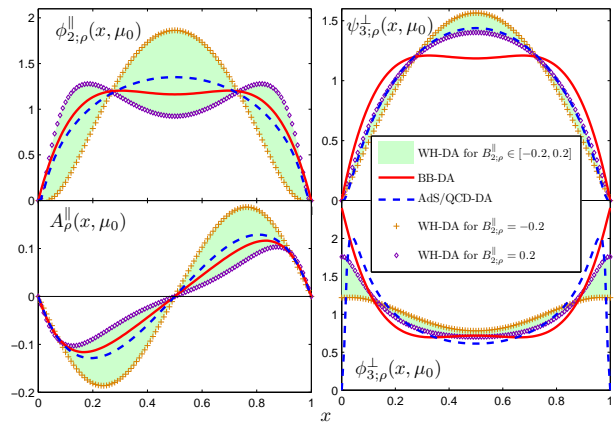}
\caption{A comparison of $\phi_{2;\rho}^\|(x,\mu_0)$, $\phi_{3;\rho}^\bot(x,\mu_0)$, $\psi_{3;\rho}^\bot(x,\mu_0)$ and $A_\rho^\|(x,\mu_0)$ under various models. The shaded band are for the WH-DA with $B^\|_{2;\rho}\in[-0.2,0.2]$. The solid and dashed lines are for BB-DA and AdS/QCD-DA, respectively. }    \label{DA:2parallel}
\end{center}
\end{figure}

In Fig.~\ref{DA:2parallel}, we present a comparison of various DA models, i.e. the WH-DA with $B^\|_{2;\rho}\in[-0.2,0.2]$, the AdS/QCD-DA and the BB-DA. Here the BB-DA stands for the usual Gegenbauer DA expansion with its second Gegenbauer moment $a_2^{\|}(1{\rm GeV})=0.15(7)$ as suggested by Ball and Braun~\cite{Ball07:rhoWF}. We also present the twist-3 DAs in Fig.~\ref{DA:2parallel}, where the WH-model is derived from the relations and the AdS/QCD and BB ones are from Ref.\cite{AdS:3} and Ref.\cite{Ball07:rhoWF}, respectively.

Fig.~\ref{DA:2parallel} shows that by varying $B^\|_{2;\rho} \in [-0.2,0.2]$, the WH-DA shall vary from the single-peaked behavior to the doubly-humped behavior. It is noted that by setting $B_{2;\rho}^{\|} =0.042$, the WH-DA behaves close to the AdS/QCD-DA; and by setting $B_{2;\rho}^\|=0.10$, the WH-DA behaves close to the BB-DA. Thus the WH-DA provides a convenient form to mimic the behaviors of various DAs suggested in the literature.

Since the LCDA $\phi_{2;\rho}^\|$ dominantly determines the $B\to\rho$ TFFs, inversely, a definite TFFs shall be helpful for fixing the properties of $\phi_{2;\rho}^\|$. In the following, we shall take the WH-DA to study the $B\to\rho$ semi-leptonic decays. To do the numerical calculation, we take $f_\rho^\|=0.216\pm0.003{\rm GeV}$~\cite{Ball07:rhoWF}, $f_B=0.160\pm0.019{\rm GeV}$~\cite{fuhb2014:Brho}, and the $b$-quark pole mass $m_b=4.80\pm0.05{\rm GeV}$. The $\rho$ and $B$-meson masses are taken as $m_\rho=0.775{\rm GeV}$ and $m_B=5.279$GeV \cite{pdg}. As a cross check of the LCSRs (\ref{TFF:A1},\ref{TFF:A2},\ref{TFF:V}), if taking the second Gegenbauer moment $a^{\perp}_2$ of $\phi_{2;\rho}^\|$ as that of Refs.\cite{Ball04:Brho,Ball1998:Brho}, we obtain consistent TFFs with those of Refs.\cite{Ball04:Brho,Ball1998:Brho} \footnote{It is noted that Refs.\cite{Ball04:Brho,Ball1998:Brho} adopt the usual correlator to derive the LCSR, in which more LCDAs that our present LCSR have to be taken into consideration.}.

\begin{table}[htb]
\begin{center}
\small
\begin{tabular}{ c   c c c }
  \hline
  ~~~$B^\|_{2;\rho}$~~~ & ~~~$-0.2$~~~ & ~~~$0.0$~~~ & ~~~$+0.2$~~~  \\
  \hline
  $s_0^{A_1}$	& 	31.5(5)	&	32.0(5)	&	32.5(5)		\\
  $M^2_{A_1}$	& 	2.4(3)	&	2.7(3)	&	3.1(3)		\\
  \hline
  $s_0^{A_2}$	& 	31.5(5)	&	32.0(5)	&	32.5(5)		\\
  $M^2_{A_2}$	& 	3.0(3)	&	2.8(3)	&	2.6(3)		\\
  \hline
  $s_0^{V}$	    & 	31.5(5)	&	32.0(5)	&	32.5(5)		\\
  $M^2_V$	    & 	3.1(3)	&	3.3(3)	&	3.5(3)		\\
  \hline
\end{tabular}
\caption{The continuum threshold $s_0$ and the Borel parameter $M^2$ under various DA models. The central values are for $m_b = 4.80$ GeV and $f_B = 0.160$ GeV.  }   \label{s0M2}
\end{center}
\end{table}

To set the parameters, such as the Borel window and the continuum threshold $s_0$, for the $B\to\rho$ TFFs' LCSR, we adopt the criteria: I) We require the continuum contribution to be less than $30\%$ of the total LCSR. II) We require all the high-twist DAs' contributions to be less than $15\%$ of the total LCSR. III) The derivatives of Eqs.(\ref{TFF:A1},\ref{TFF:A2},\ref{TFF:V}) with respect to $(-1/M^2)$ give three LCSRs for the $B$-meson mass $m_B$. And we require their predicted $B$-meson mass to be fulfilled in comparing with the experiment one, e.g. ${|m_B^{\rm th}- m_B^{\rm exp}|}/{m_B^{\rm exp}}$ less than $0.1\%$. We present the values of $s_0$ and $M^2$ for the $B\to\rho$ TFFs in Table~\ref{s0M2}.

\subsection{$B\to\rho$ TFFs and the $B\to\rho$ semi-leptonic decay}

\begin{table}[tb]
\centering
\begin{tabular}{ c  c  c  c  }
\hline
~~ ~~ & ~~~$A_1(0)$~~~ & ~~~$A_2(0)$~~~ & ~~~$V(0)$~~~ \\
\hline
~$B^\|_{2;\rho}=-0.2$~ & $0.229^{+0.049}_{-0.033}$ & $0.230^{+0.033}_{-0.024}$ & $0.253^{+0.030}_{-0.022}$\\
\hline
$B^\|_{2;\rho}=0.0$ & $0.221^{+0.035}_{-0.024}$ & $0.224^{+0.042}_{-0.029}$ & $0.258^{+0.027}_{-0.020}$\\
\hline
$B^\|_{2;\rho}=+0.2$ & $0.211^{+0.023}_{-0.017}$ & $0.220^{+0.054}_{-0.037}$ & $0.264^{+0.024}_{-0.018}$\\
\hline
\end{tabular}
\caption{The $B\to\rho$ TFFs $A_1$, $A_2$ and $V$ at the large recoil region $q^2=0$. The errors are squared averages of all the mentioned source of errors.  }\label{Endingpoint_2}
\end{table}

By using the chiral LCSRs (\ref{TFF:A1},\ref{TFF:A2},\ref{TFF:V}), we present the $B\to\rho$ TFFs under the WH-DA in Table~\ref{Endingpoint_2}, in which typical values for $B^\|_{2;\rho}=-0.2,0.0,0.2$ are adopted. The scale $\mu$ is set to be the typical momentum transfer of the process, $\mu\simeq (m_B^2-m_b^2)^{1/2}\sim 2.2{\rm GeV}$. Table~\ref{Endingpoint_2} shows that the TFFs decreases with the increment of $B^\|_{2;\rho}$ or $a^{\perp}_2$.

\begin{table*}[htb]
\begin{center}
\small
\begin{tabular}{ c  c c c  |  c c c   |   c c c  }
\hline
    &\multicolumn{3}{c|}{$B^\|_{2;\rho}=-0.2$}&\multicolumn{3}{c|}{$B^\|_{2;\rho}=0.0$}&\multicolumn{3}{c}{$B^\|_{2;\rho}=0.2$}\\ \cline{2-10}
&$A_1(0)$&$A_2(0)$&$V(0)$&$A_1(0)$&$A_2(0)$&$V(0)$&$A_1(0)$&$A_2(0)$&$V(0)$ \\ \hline
$	\psi_{3;\rho}^\bot	$&$	     /       $&$	     / $&$	0.253 	$&$	     /   $&$	     /   $&$	0.258 	$&$	     /  $&$	     /  $&$	0.264 	$	 \\
$	\phi_{3;\rho}^\bot	$&$	0.230 	     $&$	     / $&$	     /  $&$	 0.222 	 $&$	     /   $&$	     /  $&$	0.212 	$&$	     /   $&$	     /$	\\
$	A_\rho^\|	        $&$	     /	     $&$	0.221  $&$	     /  $&$	     /   $&$	0.215 	 $&$	     /  $&$	     /  $&$	0.210 	$&$	     /   $	 \\
$	C_\rho^\|	        $&$	-0.0008	     $&$	0.009  $&$	     /  $&$	 -0.0008	 $&$	0.010 	 $&$	     /  $&$	-0.0007	$&$	0.011 	$&$	     /    $	 \\
$	\Phi_{3;\rho}^\|, \widetilde{\Phi}_{3;\rho}^\|	
                        $&$	-0.00002     $&$	0.00007	$&$	     /  $&$	 -0.00002 $&$	0.00008	 $&$	     /  $&$	 -0.00002$&$	 0.00010	$&$	     /   $	 \\\hline
$	\rm{Total}	        $&$	0.229 	     $&$	0.230 	$&$	0.253 	$&$	 0.221 	 $&$	0.224 	 $&$	0.258 	$&$	0.211 	$&$	0.220 	$&$	 0.264 	 $	 \\ \hline
\end{tabular}
\caption{The $B\to\rho$ TFFs $A_1$, $A_2$ and $V$ at the large recoil region $q^2=0$, in which the WH-DA with $B^\|_{2;\rho}=-0.2$, $0.0$ and $0.2$, has been adopted. }   \label{Endingpoint_1}
\end{center}
\end{table*}

To show how the TFFs $A_1$, $A_2$ and $V$ are affected by the LCDAs, we present the TFFs at the maximum recoil region $q^2=0$ in Table~\ref{Endingpoint_1}. Up to twist-4 and $\delta^3$-order accuracy, we show that only the $\delta^1$-order LCDAs provide non-zero contributions. And among those $\delta^1$-order LCDAs, $A_{\rho}^\|$ provides the dominant contribution to $A_2$, and $\psi_{3;\rho}^\|$ provides the solitary contribution to $V$. By using the the 3-particle DAs $\Phi_{3;\rho}^\|$ and $\widetilde\Phi_{3;\rho}^\|$ suggested in Refs.\cite{Ball04:Brho,Ball1998:Brho}, we show they provide less than $0.01\%$ contributions to $A_1$ and $A_2$, which explain why those LCDAs have been neglected in many references.

\begin{table}[htb]
\begin{center}
\begin{tabular}{c c c c c }
\hline
 & ~~$F_i$~~ & ~~$a_i$~~ & ~~$b_i$~~ & ~~$\Delta$~~ \\
 \hline
              & $A_1$ & 1.036 & $-0.007$ & 0.07 \\ \cline{2-5}
$B^\|_{2;\rho}=-0.2$ & $A_2$ & 1.826 & 0.956    & 0.16   \\ \cline{2-5}
              & $V$   & 2.097 & 1.097    & 0.07 \\
\hline
             & $A_1$ & 0.949 & $-0.060$ & 0.13 \\ \cline{2-5}
$B^\|_{2;\rho}=0.0$ & $A_2$ & 1.684 & 0.748 & 0.17   \\ \cline{2-5}
             & $V$ & 1.998 & 0.969 & 0.01 \\
\hline
             & $A_1$ & 0.859 & $-0.112$ & 0.18 \\ \cline{2-5}
$B^\|_{2;\rho}=0.2$ & $A_2$ & 1.499 & 0.515 & 0.04   \\ \cline{2-5}
             & $V$ & 1.899 & 0.843 & 0.00 \\
\hline
\end{tabular}
\caption{The fitted parameters $a_i$ and $b_i$ for the $B\to\rho$ TFFs. $\Delta$ is a measure of the quality of extrapolation. } \label{analytic}
\end{center}
\end{table}

As a further step, we can apply the $B\to\rho$ TFFs to study the properties of the semileptonic decay $B\to\rho l\nu$. The LCSRs is applicable in the region $0\leq q^2\leq 14{\rm GeV^2}$, while the physical allowable region for the TFFs is $(m_B-m_\rho)^2 \sim 20{\rm GeV^2}$. Thus, certain extrapolation is needed. We adopt the formula $F_i(q^2) = F_i(0)/[1 - a_i q^2/m_B^2 + b_i(q^2/m_B^2)^2]$ to do the extrapolation, in which $F_{1,2,3}$ stand for $A_1$, $A_2$ and $V$, respectively. The parameters $a_i$ and $b_i$ are determined by requiring the ``quality'' $\Delta < 1$, which is defined as~\cite{Ball05}
\begin{equation}
\Delta=100\frac{\sum_t\left|F_i(t)-F_i^{\rm fit}(t)\right|} {\sum_t\left|F_i(t)\right|}, \label{delta}
\end{equation}
where $t\in[0,\frac{1}{2},\cdots,\frac{27}{2},14]{\rm GeV}^2$. The fitted parameters for the extrapolation are put in Table \ref{analytic}.

\begin{figure}[tb]
\begin{center}
\includegraphics[width=0.45\textwidth]{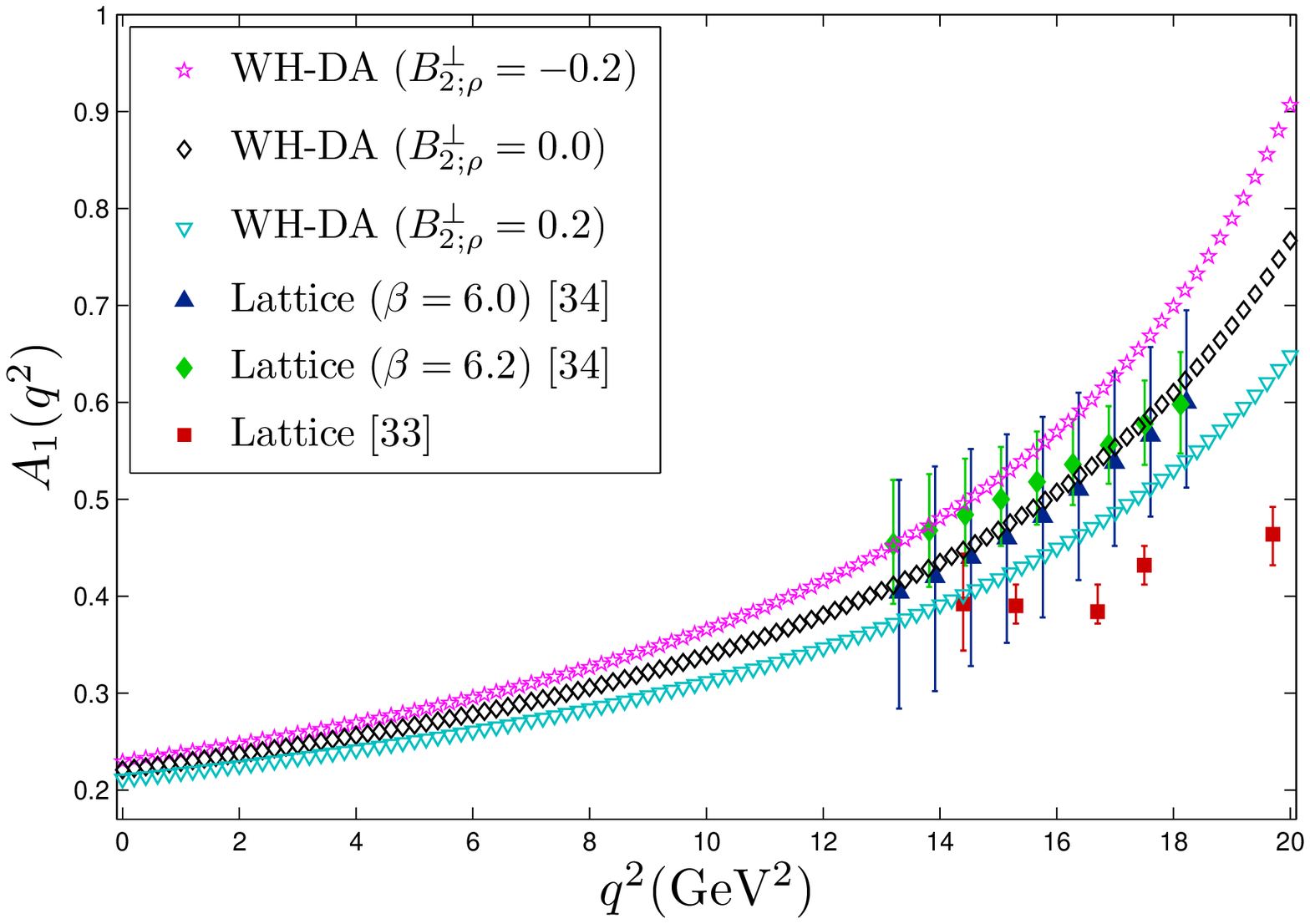}
\includegraphics[width=0.45\textwidth]{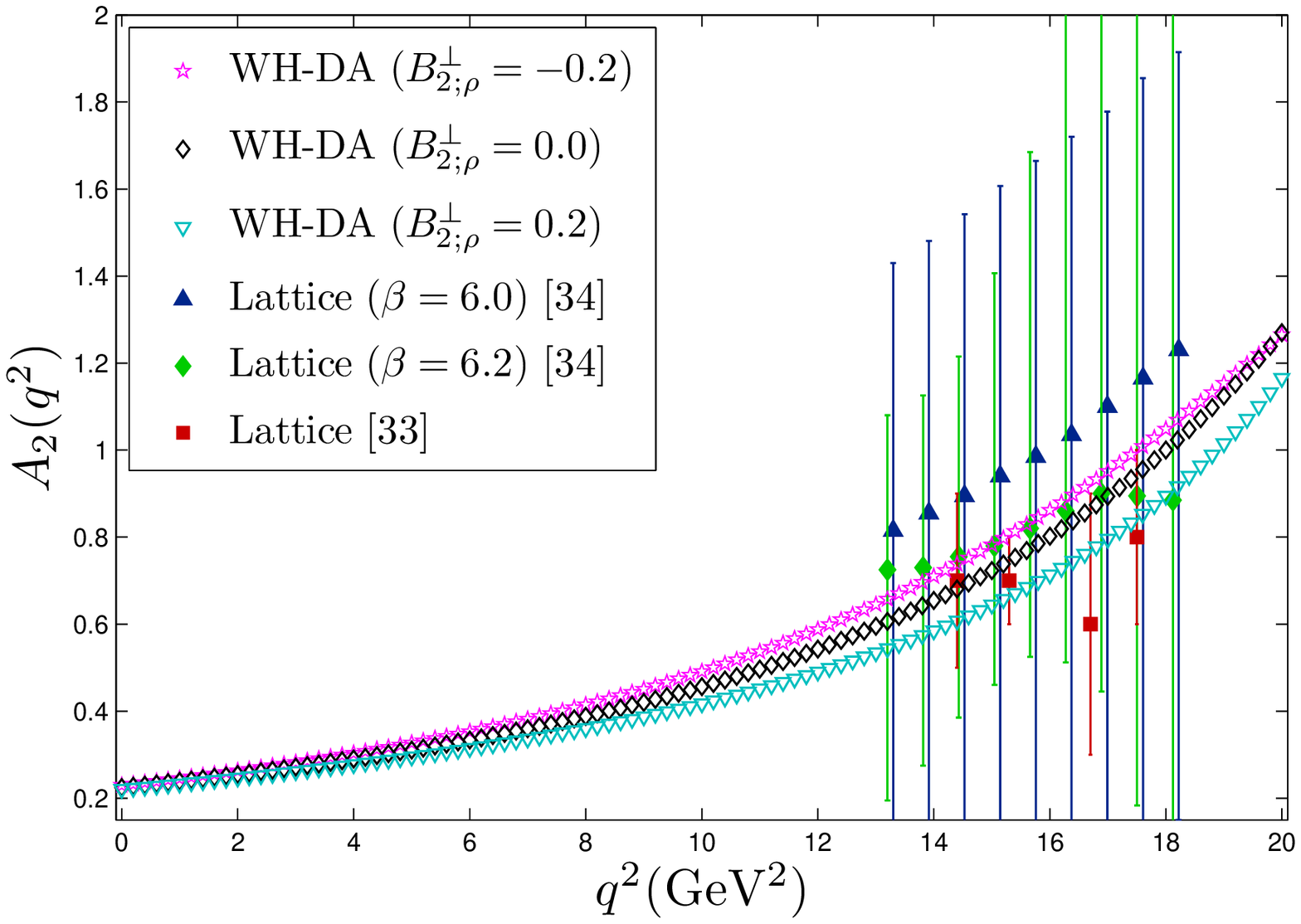}
\includegraphics[width=0.45\textwidth]{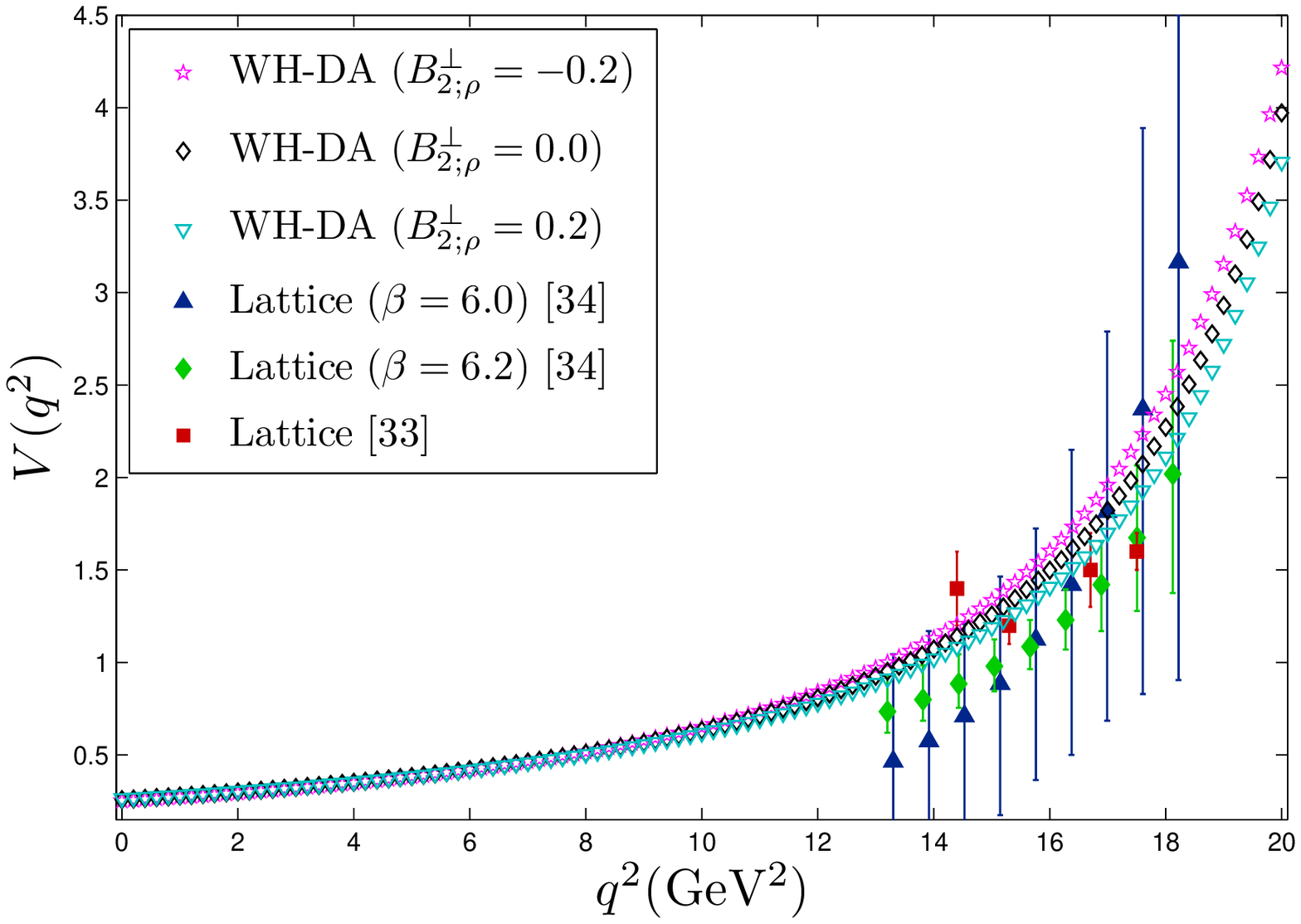}
\caption{The extrapolated $A_1(q^2)$, $A_2(q^2)$ and $V(q^2)$ under the WH-DA model, where the lattice QCD estimations~\cite{Lattice96:1,Lattice04} are included as a comparison. } \label{TFF:A1A2V}
\end{center}
\end{figure}

\begin{figure}[tb]
\begin{center}
\includegraphics[width=0.45\textwidth]{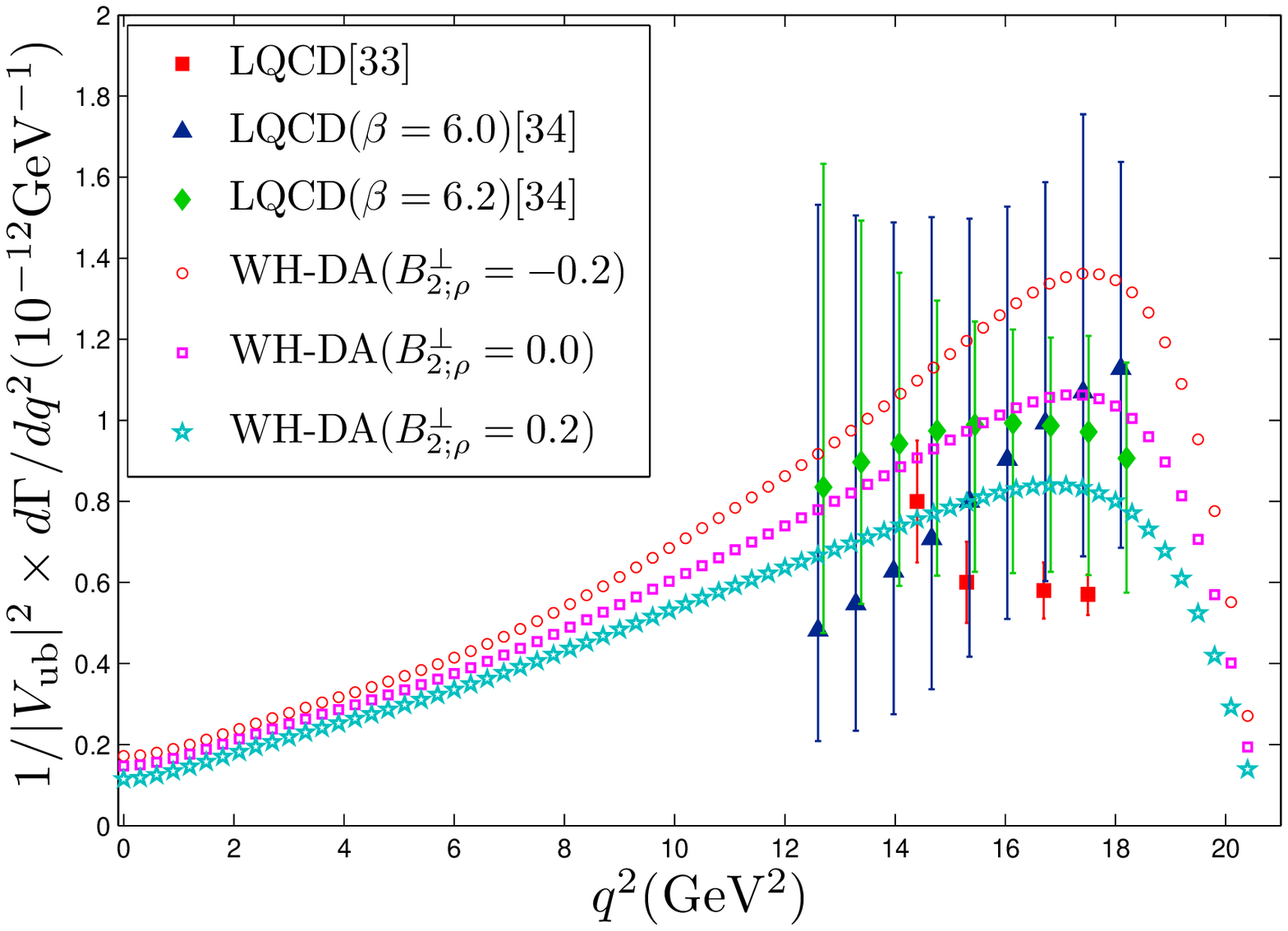}
\caption{Differential decay width $1/|V_{\rm ub}|^2 \times d\Gamma/dq^2$ for the WH-DA model. The lattice QCD estimations~\cite{Lattice96:1,Lattice04} are included as a comparison. } \label{dGamma}
\end{center}
\end{figure}

We put the extrapolated $B\to\rho$ TFFs $A_1(q^2)$, $A_2(q^2)$ and $V(q^2)$ in Fig.~\ref{TFF:A1A2V}, where the lattice QCD predictions~\cite{Lattice96:1,Lattice04} are presented for a comparison. It shows that by varying $B^\|_{2;\rho}\in[-0.2,0.2]$, the LCSRs predictions agree with the lattice QCD predictions and it shall increase with the increment of $B^\|_{2;\rho}$. One can get a strong constraint on the $\rho$-meson DAs when more precise lattice QCD results are given.

\begin{table}[htb]
\begin{center}
\small
\begin{tabular}{ccccc}
\hline
& $B^\|_{2;\rho}=-0.2$ & $B^\|_{2;\rho}=0.0$ & $B^\|_{2;\rho}=0.1$ & $B^\|_{2;\rho}=0.2$ \\
\hline
$\Gamma/|V_{\rm ub}|^2$ & $14.69^{+10.34}_{-5.57}$ & $12.20^{+6.41}_{-4.04}$ & $10.99^{+4.92}_{-3.39}$ & $10.15^{+4.05}_{-3.02}$\\
$\Gamma^\|/\Gamma^\bot$ & $0.643^{+0.409}_{-0.375}$ & $0.618^{+0.348}_{-0.363}$ & $0.597^{+0.322}_{-0.360}$ & $0.595^{+0.317}_{-0.371}$\\
\hline
\end{tabular}
\caption{Total decay width ${\Gamma}/{|V_{\rm ub}|^2}$ and the ratio $\Gamma^\|/\Gamma^\bot$ under the WH-DA model. }  \label{Gammatotal}
\end{center}
\end{table}

The differential decay width for the semileptonic decay $B\to\rho l\nu$ can be found in Ref.\cite{Ball1997:Brho}. One can cut off the uncertainty from $|V_{\rm ub}|$ by calculating the differential decay width $1/|V_{\rm ub}|^2 \times d\Gamma/dq^2$, which is presented in Fig.~\ref{dGamma}. Moreover, the total decay width ($\Gamma$) can be separated as $\Gamma^\|+\Gamma^\bot$, where $\Gamma^\|$ ($\Gamma^\bot$) stands for the decay width of the $\rho$-meson longitudinal (transverse) components. We present the total decay width ${\Gamma}$ and the ratio $\Gamma^\|/\Gamma^\bot$ in Table~\ref{Gammatotal}, where the errors in Table~\ref{Gammatotal} are squared average of the mentioned source of errors.

\subsection{The CKM matrix element $|V_{\rm ub}|$}

\begin{table}[htb]
\begin{center}
\begin{tabular}{c  c  c }
\hline
\multicolumn{2}{c}{}               & ~~~$|V_{\rm ub}|$~~~ \\ \hline
\multicolumn{2}{c}{$B^\|_{2;\rho}=-0.20$}  &  ~~ $2.51 \pm 0.38$ ~~ \\
\multicolumn{2}{c}{$B^\|_{2;\rho}=~~0.00$}  &   $2.76 \pm 0.36$ \\
\multicolumn{2}{c}{$B^\|_{2;\rho}=+0.10$}  &  $2.91 \pm 0.35$ \\
\multicolumn{2}{c}{$B^\|_{2;\rho}=+0.20$}  &  $3.02 \pm 0.35$ \\ \hline
                                                    & LCSR~\cite{Ball04:Brho}   & $2.75\pm0.24$ \\
\raisebox {1.5ex}[0pt]{BABAR~\cite{BABAR:2010}}     & ISGW~\cite{ISGR}          & $2.83\pm0.24$ \\ \hline
                                                    & LCSR~\cite{Ball04:Brho}  & $2.85\pm0.40$ \\
\raisebox {1.5ex}[0pt]{BABAR~\cite{BABAR:2005}}     & ISGW~\cite{ISGR}         & $2.91\pm0.40$ \\
\hline
\end{tabular}
\caption{The weighted average of $|V_{\rm ub}|$ in unit $10^{-3}$ from both the $B^+$-type and $B^0$-type and for the WH-DA with $B^\|_{2;\rho}=-0.2$, 0.0, 0.1 and 0.20, respectively. The estimations of the BABAR collaboration~\cite{BABAR:2010,BABAR:2005} are also presented as a comparison.}\label{Gammatota3}
\end{center}
\end{table}

We take two types of semi-leptonic decays as a try to determine the CKM matrix element $|V_{\rm ub}|$. One is the $B^0$-type via the process $B^0\to\rho^-\ell^+\nu_\ell$ with branching ratio and lifetime ${\cal B}(B^0\to\rho^-\ell^+\nu_\ell) =(2.34\pm 0.28)\times 10^{-4}$ and $\tau(B^0)= 1.519\pm 0.007$ps~\cite{pdg}. Another is the $B^+$-type via the process $B^+\to\rho^0\ell^+\nu_\ell$ with branching ratio and lifetime ${\cal B}(B^+\to\rho^0\ell^+\nu_\ell)= (1.07\pm 0.13)\times 10^{-4}$ and $\tau(B^+)= 1.641\pm 0.008$ps~\cite{pdg}. Our predicted weighted averages of $|V_{\rm ub}|$ are given in Table \ref{Gammatota3}, the BABAR predictions are also included as a comparison. The errors come from the choices of the $b$-quark mass, the Borel window and the threshold parameter $s_0$, the experimental uncertainties are from the measured lifetimes and decay ratios. Table \ref{Gammatota3} show that $|V_{\rm ub}|$ increases with the increment of $B^{\|}_{2;\rho}$. We observe that by choosing $B^\|_{2;\rho}\in[0.00,0.20]$, its central value is consistent with the BABAR prediction~\cite{BABAR:2010,BABAR:2005}. The BABAR predictions are based on the LCSR~\cite{Ball04:Brho} or ISGW~\cite{ISGR} predictions on $B\to\rho$ TFFs, respectively.

\section{Summary}

We have constructed a convenient model (\ref{DA:WH}) for the longitudinal leading-twist LCDA $\phi^\|_{2;\rho}$, in which a single parameter $B^\|_{2;\rho} \sim a^{\perp}_2$ controls its longitudinal behavior. By varying $B^\|_{2;\rho}\in[-0.20,0.20]$, the $\phi^\|_{2;\rho}$ shall evolve from a single-peaked behavior to a doubly-humped behavior. We have discussed its properties via the $B\to\rho$ TFFs by using the LCSR approach. Those chiral LCSRs for the TFFs shall be dominated (over $99\%$) by $\phi_{2;\rho}^\|$ either directly or indirectly, then those TFFs do provide us a platform to test the properties of the leading-twist $\phi_{2;\rho}^\|$. After extrapolating the TFFs to their allowable region, we compare them to those of lattice QCD predictions. Those TFFs become smaller with the increment of $B^\|_{2;\rho}$, and either a too smaller or a too larger $B^\|_{2;\rho}$ is not allowed by the lattice QCD estimations. If we have a precise lattice QCD estimation, we can get a more strong constraint on the $\rho$-meson DA behavior. In comparison to the BABAR prediction on the differential decay width $1/|V_{\rm ub}|^2 \times d\Gamma/dq^2$~\cite{BABAR:2010} and also the TFFs from Lattice QCD, we find that smaller $B^\|_{2;\rho}\lesssim-0.20$ (or $a^\|_{2;\rho}(1{\rm GeV})\lesssim-0.18$) is not suitable. As a further comparison with the BABAR prediction on the $|V_{\rm ub}|$, we observe that by choosing $B^\|_{2;\rho}\in[0.00,0.20]$, its central values show a better agreement with the BABAR prediction~\cite{BABAR:2010,BABAR:2005}. Thus, we can predict that $\phi^\|_{2;\rho}$ prefers a doubly-humped behavior other than the sing-peaked behavior of the transverse leading-twist LCDA $\phi^\perp_{2;\rho}$ as suggested by Ref.\cite{fuhb2014:Brho}.

As a final remark, the present approach can also be applied to study other vector meson LCDAs. For example, by taking the ${\rm SU}_f(3)$-breaking effect into consideration, one can further study the $K^*$ meson LCDAs. A detailed discussion on the $K^*$ meson LCDAs under the same approach of the present paper is in preparation.  \\

{\bf Acknowledgments}:  This work was supported in part by Natural Science Foundation of China under Grant No.11075225 and No.11275280, and by the Fundamental Research Funds for the Central Universities under Grant No.CQDXWL-2012-Z002.

\end{document}